\documentstyle[12pt]{article}
\newfont{\bbd}{msbm10 scaled\magstep1}

\begin{document}
\thispagestyle{empty}

\def\ve#1{\mid #1\rangle}
\def\vc#1{\langle #1\mid}

\newcommand{\p}[1]{(\ref{#1})}
\newcommand{\be}{\begin{equation}}
\newcommand{\ee}{\end{equation}}
\newcommand{\sect}[1]{\setcounter{equation}{0}\section{#1}}

\newcommand{\vs}[1]{\rule[- #1 mm]{0mm}{#1 mm}}
\newcommand{\hs}[1]{\hspace{#1mm}}
\newcommand{\mb}[1]{\hs{5}\mbox{#1}\hs{5}}
\newcommand{\Db}{{\overline D}}
\newcommand{\bea}{\begin{eqnarray}}

\newcommand{\eea}{\end{eqnarray}}
\newcommand{\wt}[1]{\widetilde{#1}}
\newcommand{\und}[1]{\underline{#1}}
\newcommand{\ov}[1]{\overline{#1}}
\newcommand{\sm}[2]{\frac{\mbox{\footnotesize #1}\vs{-2}}
           {\vs{-2}\mbox{\footnotesize #2}}}
\newcommand{\prt}{\partial}
\newcommand{\eps}{\epsilon}

\newcommand{\R}{\mbox{\rule{0.2mm}{2.8mm}\hspace{-1.5mm} R}}
\newcommand{\Z}{Z\hspace{-2mm}Z}

\newcommand{\cd}{{\cal D}}
\newcommand{\cg}{{\cal G}}
\newcommand{\ck}{{\cal K}}
\newcommand{\cw}{{\cal W}}

\newcommand{\vj}{\vec{J}}
\newcommand{\vl}{\vec{\lambda}}
\newcommand{\vz}{\vec{\sigma}}
\newcommand{\vt}{\vec{\tau}}
\newcommand{\vw}{\vec{W}}
\newcommand{\poiss}{\stackrel{\otimes}{,}}

\def\l#1#2{\raisebox{.2ex}{$\displaystyle
  \mathop{#1}^{{\scriptstyle #2}\rightarrow}$}}
\def\r#1#2{\raisebox{.2ex}{$\displaystyle
 \mathop{#1}^{\leftarrow {\scriptstyle #2}}$}}



\renewcommand{\thefootnote}{\fnsymbol{footnote}}
\newpage
\setcounter{page}{0}
\pagestyle{empty}
\begin{flushright}
{July 1999}\\
{ITP-UH-14/99}\\
{JINR E2-99-216}\\
{solv-int/9907021}
\end{flushright}
\vfill

\begin{center}
{\LARGE {\bf Supersymmetric KP hierarchy}}\\[0.3cm]
{\LARGE {\bf in N=1 superspace and its N=2 reductions}}\\[1cm]

{\large O. Lechtenfeld$^{a,1}$ and A. Sorin$^{b,2}$}
{}~\\
\quad \\
{\em {~$~^{(a)}$ Institut f\"ur Theoretische Physik, Universit\"at
Hannover,}}\\
{\em Appelstra{\ss}e 2, D-30167 Hannover, Germany}\\[10pt]
{\em {~$~^{(b)}$ Bogoliubov Laboratory of Theoretical Physics, JINR,}}\\
{\em 141980 Dubna, Moscow Region, Russia}~\quad\\

\end{center}

\vfill

\centerline{{\bf Abstract}}
\noindent
A wide class of $N{=}2$ reductions of the supersymmetric
KP hierarchy in $N{=}1$ superspace is described.
This class includes a new $N{=}2$ supersymmetric generalization
of the Toda chain hierarchy. The Lax pair representations of the bosonic
and fermionic flows, local and nonlocal Hamiltonians, finite and infinite
discrete symmetries, first two Hamiltonian structures and the recursion
operator of this hierarchy are constructed. Its secondary reduction to
new $N{=}2$ supersymmetric modified KdV hierarchy is discussed.

{}~



\vfill
{\em E-Mail:\\
1) lechtenf@itp.uni-hannover.de\\
2) sorin@thsun1.jinr.ru }
\newpage
\pagestyle{plain}
\renewcommand{\thefootnote}{\arabic{footnote}}
\setcounter{footnote}{0}

\section{Introduction}

Since the first $N=1$ supersymmetric generalization of the
bosonic KP hierarchy ---the Manin-Radul $N=1$ supersymmetric KP hierarchy
\cite{mr}--- and its reduction ---the Manin-Radul $N=1$ supersymmetric KdV
hierarchy--- appeared this subject has attracted permanent attention both
for purely academic reasons and because of various applications. Let
us mention, for example, the problem of finding a
supersymmetric hierarchy relevant for the longstanding yet unsolved
problem of constructing supermatrix models which differ non-trivially
from bosonic ones. During past years several generalizations of the
Manin-Radul supersymmetric KP hierarchy were proposed. They possess
a new type of fermionic flows \cite{m,r}, an enlarged number of bosonic
and fermionic flows \cite{tak}, or additional supersymmetries (see,
e.g. the recent paper \cite{dgs} and references therein). Recently,
a large class of new reductions of the Manin-Radul $N=1$ KP hierarchy
was discussed in the important work \cite{anp1} where bosonic and
fermionic flows respecting the original algebraic structure were
constructed. Even more recently \cite{dgs}, an $N=4$ supersymmetric KP
hierarchy was proposed and a wide class of its reductions was described
in the Lax-pair framework. It is remarkable that the supersymmetric KP
hierarchy in $N=2$ superspace actually displays an $N=4$ supersymmetry.
This doubling of supersymmetry also occurs in $N=1$ superspace where the
supersymmetric KP hierarchy is actually $N=2$ supersymmetric \cite{tak}.
It is an interesting question to find different reductions of the
latter hierarchy which preserve its $N=2$ algebraic structure. This
is the main goal of the present paper.

Using a dressing formalism, we describe a wide class of
$N=2$ reductions of the supersymmetric KP hierarchy in $N=1$
superspace. One such reduction is considered in considerable detail:
we derive its bosonic and fermionic flows, Hamiltonians, Hamiltonian
structures, recursion operator, finite and infinite discrete symmetries
and its reduction to a new $N=2$ supersymmetric modified KdV
hierarchy. As a byproduct we obtain a new version of the $N=2$
supersymmetric Toda chain equation which is related with the infinite
discrete symmetries of the reduced $N=2$ KP hierarchy.

We point out that our construction of a new class of
$N=2$ supersymmetric integrable hierarchies in $N=1$ superspace
builds upon some recent results on supersymmetric hierarchies
\cite{tak,ls,anp1,ols,dgs}. Let us describe the content of this paper.
In section 2 we review the supersymmetric KP hierarchy
in $N=1$ superspace within the framework of the dressing approach and
demonstrate that it is $N=2$ supersymmetric. In section 3 we
discuss a consistent reduction of the $N=2$ supersymmetric KP hierarchy
preserving its algebraic structure. We find its
finite and infinite discrete symmetries and use them to obtain the new
Lax operators. Finally, we construct its local and nonlocal
Hamiltonians, first two Hamiltonian structures and recursion operator.
Section 4 then describes its secondary reduction to a new version of
the $N=2$ supersymmetric modified KdV hierarchy. Section 5 presents
generalizations of the reduced $N=2$ KP hierarchy to the matrix case
and closes with some open questions. An appendix contains the new version
of the $N=2$ supersymmetric Toda chain equation, includes its
zero-curvature representation and explains the origin of the Lax
operators of section 3.

\section{N=2 supersymmetric KP hierarchy}
In this section we discuss the hierarchy which is usually called the
Manin-Radul $N=1$ supersymmetric KP hierarchy and focus on the
remarkable fact that it actually possesses an $N=2$ supersymmetry.
This section is essentially based on the results obtained in \cite{tak}.

Our starting object is the $N=1$ supersymmetric dressing operator $W$
\begin{eqnarray}
W \equiv 1+\sum_{n=1}^\infty ~(w^{(b)}_n + w^{(f)}_nD)~{\partial}^{-n},
\label{dreskp}
\end{eqnarray}
where the all functions $w_n\equiv w_n(Z)$ involved into the $W$ are the
$N=1$ superfields depending on coordinates $Z\equiv (z,{\theta})$.
Further, $D$ is the fermionic covariant derivative which, together with
the supersymmetry generator $Q$, form the
algebra\footnote{We explicitly present only non-zero brackets
in this paper.}
\begin{eqnarray}
\{ D,D\} = +2{\partial}, \quad \{ Q,Q\} = -2{\partial}
\label{alg00}
\end{eqnarray}
with the standard superspace realization:
\begin{eqnarray}
D\equiv \frac{\partial}{\partial {\theta}} +
{\theta} {\partial}, \quad
Q\equiv \frac{\partial}{\partial {\theta}} -
{\theta} {\partial}.
\label{algDQ0}
\end{eqnarray}

Our aim now is to construct a maximal set of consistent Sato
equations for the dressing operator $W$ which represent the flows
of the extended supersymmetric KP hierarchy in $N=1$ superspace.

We begin by choosing a basis $\{D,~Q, ~[{\theta},D], ~{\partial}\}$
of first-order differential operators which are point-wise $(z)$
linearly independent. We then consider arbitrary powers of these and dress
them by the dressing operator $W$, obtaining the operators
\begin{eqnarray}
L_{l} \equiv W{D}^{l}W^{-1}, \quad
M_{l} \equiv W{Q}^{l}W^{-1}, \quad N_{l}\equiv
W \frac{1}{2}[{\theta},~D]{\partial}^{l}W^{-1}
\label{laxkp}
\end{eqnarray}
with the obvious properties:
\begin{eqnarray}
L_{l} \equiv (L_{1})^{l}, \quad
M_{l} \equiv (M_{1})^{l}, \quad
L_{2l} \equiv (-1)^{l}M_{2l}=W{\partial}^{l}W^{-1}.
\label{prop}
\end{eqnarray}

Using the operators \p{laxkp} we construct
consistent Sato equations for $W$,
\begin{eqnarray}
&& {\textstyle{\partial\over\partial t_l}}W= -(L_{2l})_-W, \quad
\quad ~~U_{l}W= -(N_{l})_- W,\nonumber\\
&&D_l W= -(L_{2l-1})_- W,\quad ~~Q_lW= -(M_{2l-1})_- W,
\label{satokp}
\end{eqnarray}
where the subscript $-$ ( $+$ ) denotes the purely pseudo-differential
(differential) part of an operator.
The bosonic (fermionic) evolution derivatives
$\{{\textstyle{\partial\over\partial t_l}}, ~U_l\}$
(~$\{ D_l, ~Q_l\}$~) generating
bosonic (fermionic) flows of the hierarchy under consideration
have the following length dimensions:
\begin{eqnarray}
[{\textstyle{\partial\over\partial t_l}}]=[U_l]=-l, \quad
[D_l]=[Q_l]=-l+\frac{1}{2}.
\label{dimtimes}
\end{eqnarray}

We would like to recall that the subset of flows
$\{{\textstyle{\partial\over\partial t_l}}, ~D_l\}$ by itself forms
a hierarchy which is usually called the Manin-Radul $N=1$ supersymmetric
KP hierarchy \cite{mr}. The extra flows $\{U_l, ~Q_l\}$, when added to
the $N=1$ KP hierarchy, produce an extended hierarchy possessing a richer
algebraic structure. This extended supersymmetric
hierarchy was called the maximal SKP hierarchy in \cite{tak}.

In order to calculate the flow algebra of the extended hierarchy, one
can use a supersymmetric generalization \cite{stan,tak} of the Radul
map \cite{rad} which is a homomorphism between the flow algebra
we are looking for and the algebra of the operators $L_{l}$,
$M_{l}$ and $N_{l}$ \p{laxkp}. The resulting nonzero brackets are
\begin{eqnarray}
\quad \quad \Bigl\{D_k\,,\,D_l\Bigr\}=
-2\;\frac{{\partial}}{{\partial t_{k+l-1}}}, \quad
\Bigl\{Q_k\,,\,Q_l\Bigr\}=
+2\;\frac{{\partial}}{{\partial t_{k+l-1}}},
\label{alg1}
\end{eqnarray}
\begin{eqnarray}
&&\Bigl[U_k\,,\,D_l\Bigr]=Q_{k+l}, \quad
~\Bigl[U_k\,,\,Q_l\Bigr]=D_{k+l}.
\label{algqqbar}
\end{eqnarray}
The algebra (\ref{alg1}--\ref{algqqbar})
may be realized in the superspace $\{t_k,\theta_k,\rho_k, h_k \}$,
\begin{eqnarray}
&& D_k= \frac{\partial}{\partial \theta_k}- \sum^{\infty}_{l=1}\theta_l
\frac{\partial}{{\partial t_{k+l-1}}},\quad
Q_k=\frac{\partial}{\partial {\rho}_k}+
\sum^{\infty}_{l=1}{\rho}_l
\frac{\partial}{{\partial t_{k+l-1}}}, \nonumber\\
&& \quad \quad \quad
U_k=\frac{\partial}{\partial h_k}-
\sum^{\infty}_{l=1}({\theta}_l
\frac{\partial}{{\partial {\rho}_{k+l}}}+{\rho}_l
\frac{\partial}{{\partial {\theta}_{k+l}}}),
\label{covder}
\end{eqnarray}
where $t_k, h_k$ ($\theta_k,\rho_k$) are bosonic (fermionic) abelian
evolution times with length dimensions
\begin{eqnarray}
[t_k]=[h_k]=k, \quad [\theta_k] =[\rho_k]=k-\frac{1}{2}.
\label{dim}
\end{eqnarray}

A simple inspection of the superalgebra (\ref{alg1}--\ref{algqqbar})
shows that the flows ${\textstyle{\partial\over\partial t_1}}$,
$U_0$, $D_1$ and $Q_1$ form a finite-dimensional subalgebra which
is isomorphic to the well-known $N=2$ supersymmetry algebra including its
$gl(1)$ automorphism. For this reason the maximal SKP hierarchy may also
be called the $N=2$ supersymmetric KP hierarchy.

It is instructive to introduce a new basis,
\begin{eqnarray}
\{ D,Q, D_l,Q_l\} \Longrightarrow
\{{\cal D},{\overline {\cal D}},{\cal D}_l,{\overline {\cal D}}_l\},
\label{N2basis}
\end{eqnarray}
\begin{eqnarray}
{\cal D}\equiv \frac{1}{\sqrt{2}}(Q+D), && \quad
{\overline {\cal D}}\equiv \frac{1}{\sqrt{2}}(Q-D), \nonumber\\
{\cal D}_k\equiv \frac{1}{\sqrt{2}}((-1)^{l-1}Q_k+ D_k), && \quad
{\overline {\cal D}}_k\equiv \frac{1}{\sqrt{2}}
((-1)^{l-1}Q_k-D_k)
\label{trans}
\end{eqnarray}
in which the algebras (\ref{alg00}) and (\ref{alg1}--\ref{algqqbar})
read
\begin{eqnarray}
\{{\cal D},{\overline {\cal D}}\}=-2{\partial}, \quad
\Bigl\{{\cal D}_k\,,\,{\overline {\cal D}}_l\Bigr\}=
+2\;\frac{{\partial}}{{\partial t_{k+l-1}}}, \quad
\Bigl[U_k\,,\,{\cal D}_l\Bigr]={\cal D}_{k+l}, \quad
~\Bigl[U_k\,,\,{\overline {\cal D}}_{l}\Bigr]=-{\overline {\cal D}}_{k+l}.
\label{N2alg0}
\end{eqnarray}
In this basis the flows \p{satokp} take the form
\begin{eqnarray}
&& {\textstyle{\partial\over\partial t_l}}W= -
(W{\partial}^{l}W^{-1})_-W,
\quad \quad ~~U_{l}W= -(W{\theta}\frac{\partial}{\partial {\theta}}
{\partial}^{l}W^{-1})_- W,\nonumber\\
&&{\cal D}_l W= -(W{\cal D}{\partial}^{l-1}W^{-1})_-W,
\quad {\overline {\cal D}}_l W= -(W
{\overline {\cal D}}{\partial}^{l-1}W^{-1})_-W,
\label{satokpn2}
\end{eqnarray}
and one can easily recognize that the subflows
$\{{\textstyle{\partial\over\partial t_l}}, ~{\cal D}_l\}$ form
the Mulase-Rabin $N=1$ supersymmetric KP hierarchy \cite{m,r}.
As we have already seen earlier, the maximal SKP hierarchy
includes the Manin-Radul $N=1$ supersymmetric hierarchy as well.
Therefore, we come to the conclusion that it actually comprises
both the Manin-Radul and Mulase-Rabin $N=1$ supersymmetric KP hierarchies.

It is easy to see that the flows \p{satokpn2} are
form--invariant with respect to the $GL(1)$ automorphism transformation
of the $N=2$ supersymmetry algebra (\ref{N2alg0}),
\begin{eqnarray}
&& (\frac{{\partial}}{{\partial t_l}},\quad U_l) \quad  \Longrightarrow
\quad \quad \quad \quad \quad
(\frac{{\partial}}{{\partial t_l}},\quad U_l), \nonumber\\
&&({\cal D}, \quad {\cal D}_l)\quad ~\Longrightarrow
\quad \exp{(+i\phi)}~({\cal D}, \quad {\cal D}_l ), \nonumber\\
&&({\overline {\cal D}},\quad {\overline {\cal D}}_l) \quad
~\Longrightarrow \quad \exp{(-i\phi)}~({\overline {\cal D}}, \quad
{\overline {\cal D}}_l),
\label{N2u1}
\end{eqnarray}
where $\phi$ is an arbitrary parameter. Nevertheless, it is a very
non-trivial task to find a realization of these transformations for the
superfunctions $w_n^{f}$ and $w_n^{b}$ involved in the dressing operator
$W$ \p{dreskp}. We will return to discuss this point for
the case of the reduced $N=2$ KP hierarchy (see paragraph after eqs.
\p{recrelN2basis}). Let us finally emphasize that the $GL(1)$ covariance
of the flows was hidden in the basis \p{algDQ0}, \p{covder}, while it
becomes manifest in the new basis \p{trans}.

\section{Reduction of the N=2 KP hierarchy}
\subsection{Bosonic and fermionic flows}
In this subsection we consider a reduction of the $N=2$ supersymmetric
KP hierarchy which preserves its flow algebra
(\ref{alg1}--\ref{algqqbar}).

Let us introduce the following constraint on the operator
$M_{1}$ \p{laxkp}:
\begin{eqnarray}
M_{1} = {\cal M} \equiv Q + v D^{-1} u
\label{def1}
\end{eqnarray}
(its nature is explained in the appendix).
The operator ${\cal M}$ possesses the following important
property\footnote{Let us recall the
operator conjugation rules: $D^{T}=-D$,
$(OP)^{T}=(-1)^{d_Od_P}P^{T}O^{T}$, where $O$ ($P$) is an arbitrary
operator with the Grassmann parity $d_O$ ($d_P$), and $d_O=0$
($d_O=1$) for bosonic (fermionic) operators $O$. All other rules can be
derived using these. Hereafter, we use the notation $(Of)$ for
an operator $O$ acting only on a function $f$ inside the brackets.}:
\begin{eqnarray}
({\cal M}^{l})_-= \sum_{k=0}^{l-1}
({\cal M}^{l-k-1}v)D^{-1}(({\cal M}^{k})^{T}u), \quad l=0,1,2...
\quad,
\label{ar}
\end{eqnarray}
which can be proved by induction similarly to analogous formula in
the bosonic case \cite{eor}. Equation \p{ar} coincides with
formula of ref. \cite{anp1} where other reductions of the Manin-Radul
supersymmetric hierarchy \cite{mr} were discussed.

Substituting the expression \p{laxkp} for $M_{1}$ in
terms of the dressing operator $W$ \p{dreskp} into the constraint
\p{def1}, the latter becomes
\begin{eqnarray}
WQW^{-1} = Q + v D^{-1} u
\label{def1eq}
\end{eqnarray}
and gives an equation for $W$ which can be solved iteratively.
The unique solution ${\cal W}$ \p{dreskp} is determined by
\begin{eqnarray}
&& \quad \quad \quad \quad \quad
w^{(f)}_1\equiv -Q^{-1}(uv), \quad
w^{(b)}_1\equiv -Q^{-1}(vDu+uvQ^{-1}(uv)), \nonumber\\
&&w^{(f)}_2\equiv -Q^{-1}(vu~'+uvDQ^{-1}(uv))+
(Q^{-1}(uv))Q^{-1}(vDu+uvQ^{-1}(uv)), \quad \ldots
~~.~~~~~~
\label{def11}
\end{eqnarray}
Replacing $W$ by ${\cal W}$ in eqs. \p{laxkp} one can obtain
the reduced operators ${\cal L}_l$ and ${\cal N}_l$ as well.
As an example, we present a few terms of the $D^{-1}$ expansion of
${\cal L}_1$,
\begin{eqnarray}
{\cal L}_{1}  \equiv  {\cal W}D{\cal W}^{-1} & = &
D+2w^{(f)}_1-(Dw^{(f)}_1)D^{-1}
-((Dw^{(b)}_1)-2w^{(f)}_2+w^{(f)}_1Dw^{(f)}_1 \nonumber\\
&+& 2w^{(f)}_1 w^{(b)}_1)D^{-2}
-((D(w^{(f)}_2-w^{(f)}_1w^{(b)}_1))-
(Dw^{(f)}_1)^{2})D^{-3}+\ldots ~, ~~~~~
\label{def2}
\end{eqnarray}
where the functions $w^{(b)}_n$ and $w^{(f)}_n$ are defined in eqs.
\p{def11}.

The most complicated task now is to construct a consistent set of Sato
equations for the reduced ${\cal W}$, generalizing the unreduced equations
\p{satokp} and preserving their algebraic structure
(\ref{alg1}--\ref{algqqbar}). Recently, a similar task was carried out in
\cite{anp1} for some reductions of the Manin-Radul
$N=1$ supersymmetric KP hierarchy \cite{mr} as well as in \cite{dgs}
for the reduced $N=4$ KP hierarchy, and we essentially use the ideas
developed there. We succeeded in this construction
only for the reduced $ {\textstyle{\partial\over\partial t_l}}$, $D_l$,
and $Q_l$ flows. Nevertheless, as will be clear in what follows, the
remaining $U_l$ flows can be restored using the zero flow $U_0$
\p{ueqs} and the recurrence relations \p{recrel} (see the paragraph after
eqs. \p{flow3}).

The resulting Sato equations have the following form:
\begin{eqnarray}
 {\textstyle{\partial\over\partial t_l}}{\cal W}=
-({\cal L}_{2l})_-{\cal W}, \quad
D_l {\cal W}= -({\cal L}_{2l-1})_-{\cal W},\quad
~Q_l{\cal W}=-(({\cal M}_{2l-1})_- - {\widetilde {\cal M}}_{2l-1})
{\cal W},
\label{satoflow}
\end{eqnarray}
where a new operator ${\widetilde {\cal M}}_{2l-1}$ has been introduced,
\begin{eqnarray}
{\widetilde {\cal M}}_{2l-1}\equiv 2\sum_{k=0}^{l-2}
({\cal M}^{2(l-k)-3}v)D^{-1}(({\cal M}^{2k+1})^{T}u),
\label{conj3}
\end{eqnarray}
which is necessary for the consistency of the equations.
The flows can easily be rewritten in Lax-pair form,
\begin{eqnarray}
&&{\textstyle{\partial\over\partial t_l}}{\cal M} =
-[({\cal L}_{2l})_{-} ,{\cal M}]=[({\cal L}_{2l})_{+} ,{\cal M}],
\nonumber\\ &&D_l{\cal M} = -\{ ({\cal L}_{2l-1})_-,{\cal M}\}=
\{({\cal L}_{2l-1})_+,{\cal M}\}, \nonumber\\
&&Q_l{\cal M} = -\{ ({\cal M}_{2l-1})_- -
{\widetilde {\cal M}}_{2l-1},{\cal M}\}= \{ ({\cal M}_{2l-1})_{+}+
{\widetilde {\cal M}}_{2l-1}, {\cal M}\} - 2{\cal M}^{2l},
\label{qfl}
\end{eqnarray}
and, with the help of equation \p{ar}, lead to the following flow
equations for the superfields $v$ and $u$:
\begin{eqnarray}
&& {\textstyle{\partial\over\partial t_l}}v = (({\cal L}_{2l})_+v), \quad
~~{\textstyle{\partial\over\partial t_l}}u = -(({\cal L}_{2l})^{T}_+u),
\nonumber\\
&&D_lv =(({\cal L}_{2l-1})_+v),
\quad D_{l}u =-(({\cal L}_{2l-1})^{T}_+u), \nonumber\\
&&Q_{l}v =((({\cal M}_{2l-1})_{+}+
{\widetilde {\cal M}}_{2l-1} - 2{\cal M}_{2l-1})v), \nonumber\\
&&Q_{l}u = -((({\cal M}_{2l-1})^{T}_{+}+
({\widetilde {\cal M}}_{2l-1} - 2{\cal M}_{2l-1})^{T})u).
\label{modflv}
\end{eqnarray}

Using eqs. \p{modflv} for the bosonic and fermionic
flows, we present the first few of them, for illustration\footnote{We have
rescaled some evolution derivatives to simplify the presentation
of some formulae.},
\begin{eqnarray}
&&\quad \quad \quad \quad {\textstyle{\partial\over\partial t_0}}
\left(\begin{array}{cc} v\\ u \end{array}\right) =
\left(\begin{array}{cc} +v\\ -u \end{array}\right), \quad
{\textstyle{\partial\over\partial t_1}}
\left(\begin{array}{cc} v\\ u \end{array}\right) =
{\partial}\left(\begin{array}{cc} v\\ u \end{array}\right),
\nonumber\\ &&{\textstyle{\partial\over\partial t_2}} v =
+v~'' -  2uv(DQv)+(DQv^2u)+v^2(DQu) -2v(uv)^2,
\nonumber\\
&&{\textstyle{\partial\over\partial t_2}} u =
-u~'' -  2uv(DQu)+(DQu^2v)+u^2(DQv) +2u(uv)^2,
\label{eqs}
\end{eqnarray}
\begin{eqnarray}
&& \quad \quad \quad \quad
D_1 v= -Dv+ 2vQ^{-1}(uv),
\quad D_1 u= -Du- 2uQ^{-1}(uv), \nonumber\\
&& \quad \quad \quad \quad
Q_1 v= -Qv- 2vD^{-1}(uv),
\quad Q_1 u= -Qu+ 2uD^{-1}(uv), \nonumber\\
&& D_2 v = -Dv~'+ 2v~'Q^{-1}(uv) + (Dv)Q^{-1}D(uv)
+ vQ^{-1}[uv~' +(Dv)(Du)],\nonumber\\
&& D_2 u =+Du~' + 2u~'Q^{-1}(uv) +
(Du)Q^{-1}D(uv) + uQ^{-1}[vu~' + (Du)(Dv)], \nonumber\\
&& Q_2 v = -Qv~'- 2v~'D^{-1}(uv) + (Qv)D^{-1}Q(uv)
- vD^{-1}[uv~' - (Qv)(Qu)],\nonumber\\
&& Q_2 u =+Qu~' - 2u~'D^{-1}(uv) +
(Qu)D^{-1}Q(uv) - uD^{-1}[vu~' - (Qu)(Qv)],
\label{ff2-}
\end{eqnarray}
\begin{eqnarray}
U_0\left(\begin{array}{cc} v\\ u \end{array}\right) =
{\theta}D \left(\begin{array}{cc} v\\ u \end{array}\right),
\label{ueqs}
\end{eqnarray}
where the zero flow $U_0$ was constructed by hand so that it
satisfies the algebra (\ref{alg1}--\ref{algqqbar}).

We would like to close this subsection with a few remarks.

First, observing eqs. \p{modflv} we learn that
the reduced $N=2$ KP flows (except for
${\textstyle{\partial\over\partial t_l}}$) are nonlocal in general
(for an example, see eqs. (\ref{ff2-}) ). This property of flows is just
the result of the reduction.

Second, the flows $\{{\textstyle{\partial\over\partial t_1}},~U_0,
~D_1,~Q_1\}$ forming the $N=2$ supersymmetry algebra are non-locally
and non-linearly realized in terms of the initial superfields $v$ and $u$.
However, there exists another superfield basis
$\{{\widehat v},{\widehat u}\}$, defined as
\begin{eqnarray}
{\widehat v}\equiv v \exp{\{+[\theta,D^{-1}](uv)\}}, \quad
{\widehat u}\equiv u\exp{\{-[\theta,D^{-1}](uv)\}},
\label{N4lintransf}
\end{eqnarray}
which localizes and linearizes the $N=2$ supersymmetry realization into
\begin{eqnarray}
{\textstyle{\partial\over\partial t_1}}={\partial},
\quad D_1 =-D, \quad Q_1 =Q, \quad U_0={\theta}D.
\label{N4linflows}
\end{eqnarray}
However, in this basis even the flows
${\textstyle{\partial\over\partial t_l}}$ for $l \geq 2$ are nonlocal.

Let us finally stress that, in distinction to all known Lax operators
used before in the supersymmetric literature,
the Lax operators proposed in this subsection do not
respect supersymmetry because they contain both the fermionic
covariant derivative $D$ and the supersymmetry generator $Q$.
Nevertheless, the resulting hierarchy is $N=2$ supersymmetric.

\subsection{\bf Discrete symmetries, Darboux-B\"acklund
transformations and solutions}
In this subsection we discuss finite and infinite discrete
symmetries of the reduced hierarchy, and use them to construct
its solutions and new Lax operators.

Direct verification shows that the flows (\ref{eqs}--\ref{ueqs})
admit the two involutions:
\begin{eqnarray}
&&(v,u)^{*}= i(u,v), \quad
(z,{\theta})^{*}=(z,{\theta}),
\quad (t_p,U_p,D_p,Q_p)^{*}=(-1)^{p-1}(t_p,-U_p,D_p,Q_p),
\label{conj*}
\end{eqnarray}
\begin{eqnarray}
&& (v,u)^{\dagger}= (u,v), \quad
(z,{{\theta}})^{\dagger}=(-z,{\theta}),
\quad (t_p,U_p,D_p,Q_p)^{\dagger}=(-t_p,U_p,Q_p,D_p),
\label{conjdagger}
\end{eqnarray}
which are consistent with their algebra (\ref{alg00}),
(\ref{alg1}--\ref{algqqbar}). A third involution can easily be
derived by multiplying these two.

It is a simple exercise to check that all the flows (\ref{qfl})
(or \p{modflv} ) also possess the involution \p{conj*}, using the
following involution property of the dressing operator ${\cal W}$:
\begin{eqnarray}
{\cal W}^{*}= ({\cal W}^{-1})^{T}
\label{Wconj*}
\end{eqnarray}
which results from eq. \p{def1eq} and its consequences
\begin{eqnarray}
({{\cal L}_l})^{*}= (-1)^{\frac{l(l+1)}{2}}({\cal L}_l)^{T}, \quad
({{\cal M}_l})^{*}= (-1)^{\frac{l(l+1)}{2}}({\cal M}_l)^{T}, \quad
({{\widetilde {\cal M}}_{2l-1}})^{*}= (-1)^{l}
({\widetilde {\cal M}}_{2l-1})^{T},
\label{opconj*}
\end{eqnarray}
for the operators entering eqs. (\ref{qfl}). As regards
to the involution (\ref{conjdagger}), we do not have a direct
proof for it due to the very complicated transformation property
of the dressing operator. However, a simple proof can be given
using the recurrence relations \p{recrel}, to be derived later.

Besides the involutions (\ref{conj*}--\ref{conjdagger}) the flows
(\ref{eqs}--\ref{ueqs}) possess an infinite-dimensional group of
discrete Darboux transformations (see eq. \p{Darboux} of appendix)
\begin{eqnarray}
&&\quad  \quad \quad \quad (v,~u)^{\ddagger}=(~v(QD\ln v-uv),~
\frac{1}{v}~),
\nonumber\\ &&(z,{{\theta}})^{\ddagger}=(z,{\theta}),\quad
(t_p,U_p,D_p,Q_p)^{\ddagger}=(t_p,U_p,-D_p,-Q_p),
\label{discrsymm}
\end{eqnarray}
\begin{eqnarray}
{\cal M}^{\ddagger} = - {\cal T}{\cal M}{\cal T}^{-1},
\quad {\cal T}\equiv v D v^{-1}.
\label{DBtransform1}
\end{eqnarray}
Let us remark that
formula \p{DBtransform1} represents the Darboux-B\"acklund
transformation\footnote{For the reduced Manin-Radul $N=1$ supersymmetric
KP hierarchy the Darboux-B\"acklund transformations were discussed in
\cite{anp1} (see also references therein).} of the Lax operator ${\cal M}$
\p{def1}.

Applying involutions (\ref{conj*}--\ref{conjdagger})
and the discrete group \p{discrsymm} to the Lax operator
${\cal M}$ \p{def1} one can derive other consistent Lax operators
\begin{eqnarray}
{\cal M}^{*}=Q - u D^{-1} v \equiv -{\cal M}^{T}, \quad
\quad {\cal M}^{\dagger} = D + u Q^{-1} v,
\label{conjlax}
\end{eqnarray}
\begin{eqnarray}
{\cal M}^{((j+1)\ddagger)} = D_- +
v^{((j+1)\ddagger)} D_{+}^{-1} u^{((j+1)\ddagger)}\equiv -
{\cal T}^{(j\ddagger)}{\cal M}^{(j\ddagger)}{{\cal T}^{(j\ddagger)}}^{-1},
\quad {\cal T}^{(j\ddagger)}\equiv v^{(j\ddagger)} D_{+}
{v^{(j\ddagger)}}^{-1}
\label{DBtransform}
\end{eqnarray}
which generate isomorphic flows.
${\cal M}^{(j\ddagger)}$ is obtained from ${\cal M}$ by applying
$j$ times the discrete transformation \p{DBtransform1}, e.g.
${\cal M}^{(3\ddagger)} \equiv
(({\cal M}^{\ddagger})^{\ddagger})^{\ddagger}$,
${\cal M}^{(0\ddagger)} \equiv {\cal M}$.

Generalizing results obtained in \cite{ols,dgs} one can construct an
infinite class of solutions for the reduced hierarchy under consideration.
We briefly present this construction and refer to \cite{ols,dgs} for
details.

The simplest solution of the hierarchy corresponds to
\begin{eqnarray}
u = 0,
\label{boncond}
\end{eqnarray}
in which case the bosonic and fermionic flows for the remaining
superfield $v\equiv -{\tau}_0$ are linear and have the following form:
\begin{eqnarray}
{\textstyle{\partial\over\partial t_k}}{\tau}_0={\partial}^{k}{\tau}_0,
\quad D_k{\tau}_0=-D {\partial}^{k-1}{\tau}_0,
\quad Q_k{\tau}_0=Q{\partial}^{k-1}{\tau}_0.
\quad U_k{\tau}_0={\theta} D{\partial}^{k}{\tau}_0.
\label{eqv}
\end{eqnarray}
To derive these equations it suffices to take into account the
length dimensions \p{dimtimes} of the evolution derivatives, their algebra
(\ref{alg1}--\ref{algqqbar}) and the invariance of all flows
(\ref{qfl}) with respect to the $GL(1)$ transformations
\begin{eqnarray}
(~v, u~) \quad \Longrightarrow \quad
(~\exp{(+i\beta)}~v, ~\exp{(-i\beta)}~u~)
\label{u1}
\end{eqnarray}
which is obvious due to the invariance of the reduction constraint
\p{def1eq}.

For technical reasons, we restrict the analysis of the hierarchy to
the case when only the flows ${\textstyle{\partial\over\partial t_k}}$,
$D_k$ and $Q_k$ (but not $U_k$) are
considered. Then, using the realization \p{covder}, the solution of eqs.
\p{eqv} is
\begin{eqnarray}
\tau_{0} = \int\!\! d\lambda\; d\eta\; \varphi(\lambda,\eta)\;
\exp\Bigl\{ x{\lambda} {-}\!
~{\eta} {\theta} {+}\! \sum^{\infty}_{k=1} \Bigl [t_k
{+}\! \eta ({\theta}_k-{\rho}_k){\lambda}^{-1}+
\theta ({\theta}_k+{\rho}_k)-
{\theta}_k \sum^{\infty}_{n=1} {\rho}_n
{\lambda}^{n-1} \Bigr]{\lambda}^{k}\Bigr \},
\label{gensol}
\end{eqnarray}
where $\varphi$ is an arbitrary fermionic function of the
bosonic ($\lambda$) and fermionic ($\eta$)
spectral parameters with length dimensions
\begin{eqnarray}
[\lambda]= -1, \qquad [\eta] = - \frac{1}{2}.
\label{dim1}
\end{eqnarray}
Applying the discrete group \p{discrsymm} to the solution constructed
$\{u=0,~v=-{\tau}_0\}$, an infinite
class of new solutions of the hierarchy is generated through an
obvious iterative procedure \cite{ols}
\begin{eqnarray}
&&v^{((2j+1)\ddagger)}\ =\ +(-1)^{j}\frac{\tau_{2j}}{\tau_{2j+1}}, \qquad\
\quad v^{(2(j+1)\ddagger)}\ =\
(-1)^{j}\frac{\tau_{2(j+1)}}{\tau_{2j+1}},
\nonumber\\[8pt]
&&u^{((2j+1)\ddagger)}\ =\ -(-1)^{j}\frac{\tau_{2j-1}}{\tau_{2j}}, \qquad
\quad u^{(2(j+1)\ddagger)}\ =\ (-1)^{j}\frac{\tau_{2j+1}}{\tau_{2j}},
\quad j=0,1,2, \ldots ,
\label{todasol}
\end{eqnarray}
where the $\tau_j$ are\footnote{ The superdeterminant is defined as
${\quad \rm sdet} \left(\begin{array}{cc} A & B \\ C & D
\end{array}\right)\ \equiv\
\det (A-BD^{-1}C ) (\det D)^{-1}$.}
\begin{eqnarray}
&& \tau_{2j}\quad\ =\ {\rm sdet} \biggl(\begin{array}{cc}
(-1)^{q}{\partial}^{p+q}\tau_0
& (-1)^{m}{\partial}^{p+m}Q\tau_0 \\
~~(-1)^{q}{\partial}^{k+q}D\tau_0
&(-1)^{m} {\partial}^{k+m}DQ\tau_0
\end{array}\biggr)^{0\leq p,q \leq j}_{0\leq k,m \leq j-1},
\nonumber\\[10pt]
&&\tau_{2j+1}\ =\ {\rm sdet} \biggl(\begin{array}{cc}
(-1)^{q}{\partial}^{p+q}\tau_0
& (-1)^{m}{\partial}^{p+m}Q\tau_0 \\
~~(-1)^{q}{\partial}^{k+q}D\tau_0
& (-1)^{m}{\partial}^{k+m}DQ\tau_0
\end{array}\biggr)^{0\leq p,q \leq j}_{0\leq k,m \leq j}.
\label{supdet}
\end{eqnarray}

\subsection{Hamiltonian structure}
In this subsection we construct local and nonlocal Hamiltonians,
first two Hamiltonian structures and the recursion operator
of the reduced hierarchy.

Let us first present our notations for the $N=1$ superspace measure and
delta function
\begin{eqnarray}
dZ \equiv dz d \theta, \quad
{\delta}^{N=1}(Z) \equiv \theta {\delta}(z),
\label{notat0}
\end{eqnarray}
as well as the realization of the inverse derivatives
\begin{eqnarray}
&&D^{-1}\equiv D{\partial}_{z}^{-1}, \quad
Q^{-1}\equiv -Q{\partial}_{z}^{-1}, \quad
{\partial}_{z}^{-1} \equiv \frac{1}{2}
\int_{-\infty}^{+\infty}dx{\epsilon}(z-x), \nonumber\\
&& \quad \quad \quad \quad
{\epsilon}(z-x)=-{\epsilon}(x-z)\equiv 1, \quad if \quad z>x
\label{derrealiz}
\end{eqnarray}
which we use in what follows. We also use the correspondence:
\begin{eqnarray}
{\textstyle{\partial\over\partial {\tau}^{a}_l}}
\equiv \{ {\textstyle{\partial\over\partial t_l}},U_l, D_l, Q_l\}
\quad \Leftrightarrow \quad
{\cal H}^{a}_l \equiv \{ {\cal H}^{t}_l,{\cal H}^{U}_l,
{\cal H}^{D}_l, {\cal H}^{Q}_l\}
\label{notat}
\end{eqnarray}
between the evolution derivatives
${\textstyle{\partial\over\partial {\tau}^{a}_l}}$ and Hamiltonian
densities ${\cal H}^{a}_l$, and, consequently, the latter ones have the
length dimensions
\begin{eqnarray}
[{\cal H}^{t}_l]=[{\cal H}^{U}_l]=-l, \quad [{\cal H}^{{D}}_l]=
[{\cal H}^{{Q}}_l]=-l+\frac{1}{2}.
\label{hams0}
\end{eqnarray}
Let us remark that the length dimensions of the Hamiltonian $H^{a}_l$,
\begin{eqnarray}
{H}^{a}_l \equiv \int d Z {\cal H}^{a}_l,
\label{hamsrel}
\end{eqnarray}
and its density ${\cal H}^{a}_l$ are different. They are related as
\begin{eqnarray}
[{H}^{a}_l]=[{\cal H}^{a}_l] + \frac{1}{2}
\label{hamsrel1}
\end{eqnarray}
because the length dimension of the $N=1$ superspace measure is not equal
to zero, $[dZ]=\frac{1}{2}$. Moreover, their Grassmann parities,
$d_{H^{a}}$ and $d_{{\cal H}^{a}}$, are opposite,
\begin{eqnarray}
d_{H^{a}}= d_{{\cal H}^{a}}+1,
\label{hams00}
\end{eqnarray}
due to the fermionic nature of the $N=1$ superspace measure \p{notat0}.

We define the residue of pseudo-differential operators ${\Psi}$,
generated by even powers of the Lax operator ${\cal M}$ \p{def1}, 
according to the rule:
\begin{eqnarray}
{\Psi} \equiv ...+(Q~res({\Psi}))D^{-1}+... \quad ,
\label{defres}
\end{eqnarray}
then, bosonic Hamiltonian
densities\footnote{Let us recall that Hamiltonian densities are
defined up to terms which are fermionic or bosonic total derivatives of
an arbitrary functional $f(Z)$ of the initial superfields subjected to the
constraint: $f(+\infty, \theta)-f(-\infty,\theta)=0$.}
can be defined as
\begin{eqnarray}
{\cal H}^{t}_l \equiv res({\cal M}_{2l}).
\label{res0}
\end{eqnarray}
The residue \p{defres} for pseudo-differential operators in
$N=1$ superspace is not the usual $N=1$ residue which is the
coefficient of the operator $D^{-1}$. We obtained this unusual definition
for the residue by some special reduction of the residue introduced in
\cite{dgs} which we discuss in what follows (see the paragraph after eq. 
\p{satoflowredq}). Actually, we do not have a proof that the necessary
properties of the residue indeed hold for the definition \p{defres},
but we have verified explicitly for the few first values of $l$
($l=1,2,3$) that eqs. \p{res0} do in fact generate conserved
quantities of the flows ${\textstyle{\partial\over\partial t_k}}$
\p{eqs}. 

Using formulae (\ref{defres}--\ref{res0}) and the relation
(\ref{ar}) one can derive the general formulae for the Hamiltonians
${H}^{t}_l$ in terms of the Lax operator ${\cal M}$ \p{def1}
\begin{eqnarray}
{H}^{t}_{l-1} =
\int dZ D^{-1}\sum_{k=0}^{2l-3}(-1)^{k}({\cal M}^{2l-3-k}v)
(({\cal M}^{k})^{T}u).
\label{hamsgen}
\end{eqnarray}
We present, for example, the explicit
expressions for the first few bosonic
Hamiltonians\footnote{When deriving eqs. (\ref{hamsgen}--\ref{hamsferm})
we integrated by parts and made essential use of realizations
\p{derrealiz} for the inverse derivatives and of the relationship
$Q\equiv D-2{\theta}{\partial}$. We also used the
following definition of the superspace integral:
$\int dZ f(Z)\equiv \int dz (Df)(z,0)$.},
\begin{eqnarray}
&& \quad \quad \quad \quad \quad \quad \quad
{H}^{t}_1 = \int dZ uv, \quad {H}^{t}_2 =\int dZ uv~', \nonumber\\
&&\quad \quad {H}^{t}_3 = \int
dZ ~[~uv~''-vu[u(DQv)-v(DQu)]-\frac{2}{3}(uv)^3~],
\label{hams}
\end{eqnarray}
\begin{eqnarray}
{H}^{U}_0 = \int dZ ~u {\theta} D v,
\label{hamsU}
\end{eqnarray}
and the first few fermionic ones,
\begin{eqnarray}
{H}^{D}_1= {H}^{Q}_1= \int dZ D^{-1}(uv),
\label{hamsfermnonl}
\end{eqnarray}
\begin{eqnarray}
&&{H}^{D}_2= \int dZ ~[~ vDu + uvQ^{-1}(uv)~], \quad
{H}^{Q}_2= \int dZ ~[~ vQu - uvD^{-1}(uv)~], \nonumber\\
&& \quad \quad \quad {H}^{D}_3 = \int dZ ~[~ vDu~' +
2vu~'Q^{-1}(uv)+v(Du)[\theta,D](uv)~], \nonumber\\
&&\quad \quad \quad {H}^{Q}_3 = \int dZ ~[~ vQu~' +
2vu~'D^{-1}(uv)+v(Qu)[\theta,D](uv)~].
\label{hamsferm}
\end{eqnarray}
The Hamiltonians (\ref{hamsU}--\ref{hamsferm}) were found
manually by requiring that they are conserved with respect to
the flows ${\textstyle{\partial\over\partial t_l}}$ \p{eqs}.

We should add that the Hamiltonians in
eqs. (\ref{res0}) for $l\geq 4$
are only conjectured to be conserved under the bosonic
flows ${\textstyle{\partial\over\partial t_k}}$ \p{qfl}.
This conjecture was checked explicitly for $l=1,2,3$.

It is well known that a bi-Hamiltonian system of
evolution equations can be represented as:
\begin{eqnarray}
{\textstyle{\partial\over\partial {\tau}^{a}_l}}
\left(\begin{array}{cc} v \\ u \end{array}\right) =
J_1 \left(\begin{array}{cc} {\delta}/{\delta v} \\
{\delta}/{\delta u} \end{array}\right) H^{a}_{l+1}=
J_2 \left(\begin{array}{cc} {\delta}/{\delta v} \\
{\delta}/{\delta u} \end{array}\right) H^{a}_{l},
\label{hameq}
\end{eqnarray}
where $J_1$ and $J_2$ are the first and second Hamiltonian structures.
In terms of these the Poisson brackets of the
superfields $v$ and $u$ are given by the formula:
\begin{eqnarray}
\{\left(\begin{array}{cc}
v(Z_1)\\ u(Z_1)\end{array}\right)
\stackrel{\otimes}{,}
\left(\begin{array}{cc} v(Z_2),u(Z_2)\end{array}\right)\}_l=
J_l(Z_1){\delta}^{N=1}(Z_1-Z_2).
\label{palg}
\end{eqnarray}

An important remark is in order. In $N=1$ superspace the
variational derivatives $\frac{\delta}{\delta v}$
and $\frac{\delta}{\delta u}$ are Grassmann odd
because of the definition
$\frac{\delta}{\delta v(Z_1)}v(Z_2)\equiv {\delta}^{N=1}(Z_1-Z_2)$
and the fermionic nature of the $N=1$ delta function \p{notat0}.

Using flows (\ref{eqs}--\ref{ueqs})
and Hamiltonians (\ref{hams}--\ref{hamsferm}), we have found the
Hamiltonian structures to be
\begin{eqnarray}
J_1= \left(\begin{array}{cc} 0 & 1 \\
-1 &  0\end{array}\right) \quad and \quad
J_2= \left(\begin{array}{cc} J_{11} & J_{12} \\
J_{21} &  J_{22} \end{array}\right)
\label{hamstr}
\end{eqnarray}
with
\begin{eqnarray}
J_{11} &\equiv & +vD^{-1}vQ-(Qv)D^{-1}v-
2vD^{-1}uvD^{-1} v \nonumber\\ && +vQ^{-1}vD
-(Dv)Q^{-1}v+2vQ^{-1}uvQ^{-1} v,\nonumber\\ J_{22} & \equiv &
-uQ^{-1}uD+(Du)Q^{-1}u+ 2uQ^{-1}uvQ^{-1} u \nonumber\\ && -uD^{-1}uQ
+(Qu)D^{-1}u-2uD^{-1}uvD^{-1} u, \nonumber\\ J_{12} & \equiv & -\partial +
\{Q,vD^{-1}u\}+ 2vD^{-1}uvD^{-1}u + \{D,vQ^{-1}u\} -
2vQ^{-1}uvQ^{-1}u,\nonumber\\ J_{21}& \equiv & -\partial - \{D,uQ^{-1}v\}-
2uQ^{-1}uvQ^{-1}v - \{Q,uD^{-1}v\} + 2uD^{-1}uvD^{-1}v.
\label{hamstr2}
\end{eqnarray}
We would like to note that, other than for the $N=4$ Toda chain hierarchy
\cite{dgs}, the Hamiltonian
structures (\ref{palg}--\ref{hamstr2}) are Grassmann odd due to the
bosonic character of the matrices $J_1$ and $J_2$
(\ref{hamstr}--\ref{hamstr2}) and the fermionic nature of the $N=1$ delta
function \p{notat0}. Odd Hamiltonian structures were
also used earlier in the description of some supersymmetric integrable
systems (for recent papers, see \cite{ffs,mcar,bd,s,p} and references
therein).

The second Hamiltonian structure $J_2$ \p{hamstr2}
is rather complicated, nonlinear and nonlocal. It
becomes linear and local in terms of the original
Toda-chain superfields $b$ and $f$ (see eq. \p{zerocurveqs3} of appendix)
\begin{eqnarray}
b\equiv uv, \quad f\equiv D\ln v.
\label{zerocurveqs3trans}
\end{eqnarray}
The corresponding
Hamiltonian structures $J^{(b,f)}_1$ and $J^{(b,f)}_2$ can be expressed
via $J_1$ and $J_2$ (\ref{hamstr}--\ref{hamstr2}) by the following
standard relation:
\begin{eqnarray}
\quad \quad \quad \quad \quad
J^{(b,f)}_l= FJ_lF^{T}, \quad
F\equiv \left(\begin{array}{cc} u & v \\
D\frac{1}{v} &  0 \end{array}\right),
\label{hamstr-bfbasis}
\end{eqnarray}
where $F$ is the matrix of Frechet derivatives corresponding to the
transformation $\{b,f\} \Rightarrow \{v,u\}$  \p{zerocurveqs3trans}.
One finds:
\begin{eqnarray}
&& \quad \quad \quad \quad
J^{(b,f)}_1= \left(\begin{array}{cc} 0 & D \\
D &  0 \end{array}\right), \quad
J^{(b,f)}_2= \left(\begin{array}{cc} J^{(b,f)}_{11} & J^{(b,f)}_{12} \\
J^{(b,f)}_{21} &  J^{(b,f)}_{22} \end{array}\right), \nonumber\\
J^{(b,f)}_{11} &\equiv & -{\partial} b - b {\partial}, \nonumber\\
J^{(b,f)}_{12} & \equiv & {\partial}D+Q b+Db
[\theta,D] -(Df)D, \nonumber\\
J^{(b,f)}_{21} & \equiv & -{\partial}D+bQ -[\theta,D]bD-D(Df), \nonumber\\
J^{(b,f)}_{22}& \equiv & -2QD+2b-2(Qf)-[(Df),[\theta,D]]
+2[\theta,D]b[\theta,D].
\label{hamstr2bf}
\end{eqnarray}

Using equations \p{hamsfermnonl}, \p{hameq} and \p{hamstr}, we obtain, for
example, the 0-th fermionic flow,
\begin{eqnarray}
Q_0 v=v{\theta}, \quad Q_0 u=-u{\theta}.
\label{ff0} \end{eqnarray}

Knowledge of the first and second Hamiltonian structures allows us to
construct the recursion operator of the hierarchy,
\begin{eqnarray}
R = J_2 J^{-1}_1 \equiv
\left(\begin{array}{cc}
J_{12}, & -J_{11} \\
J_{22}, &  -J_{21}
\end{array}\right), \quad
\frac{\partial}{\partial {\tau}^{a}_{l+1}}
\left(\begin{array}{cc} v\\u \end{array}\right) =
R \frac{\partial}{\partial {\tau}^{a}_{l}}
\left(\begin{array}{cc} v\\u \end{array}\right),
\label{recop0}
\end{eqnarray}
\begin{eqnarray}
{\textstyle{\partial\over\partial {\tau}^{a}_{l+1}}}
\left(\begin{array}{cc} v\\\ u \end{array}\right) =
R^{l} {\textstyle{\partial\over\partial {\tau}^{a}_{1}}}
\left(\begin{array}{cc} v\\u \end{array}\right), \quad
J_{l+1} = R^l J_1.
\label{hamstrn}
\end{eqnarray}
The Hamiltonian structures $J^{(b,f)}_1$ and $J^{(b,f)}_2$ \p{hamstr2bf}
(and, consequently, the original Hamiltonian structures $J_1$ and $J_2$
(\ref{hamstr}--\ref{hamstr2}) ) are obviously mutually compatible: a
deformation of the superfield $f$ to $f + \gamma \theta$,
where $\gamma$ is an arbitrary parameter, transforms $J^{(b,f)}_2$ into
the Hamiltonian structure defined by the algebraic sum
\begin{eqnarray}
\quad \quad \quad \quad \quad
J^{(b,f+\gamma\theta)}_2= J^{(b,f)}_2+\gamma J^{(b,f)}_1.
\label{hamstr-comp}
\end{eqnarray}
Therefore, the recursion operator $R$ \p{recop0} is
hereditary as the operator obtained from the compatible pair of the
Hamiltonian structures \cite{ff}.

Applying formulae \p{recop0} we obtain the following recurrence relations
for the flows:
\begin{eqnarray}
{\textstyle{\partial\over\partial {\tau}^{a}_{l+1}}}v=
+{\textstyle{\partial\over\partial {\tau}^{a}_{l}}}v~'
&+& (-1)^{d_{{\tau}^{a}}}vD^{-1}
{\textstyle{\partial\over\partial {\tau}^{a}_{l}}}
{\cal H}^{Q}_2 - [(Qv)+v(D^{-1}uv)]D^{-1}
{\textstyle{\partial\over\partial {\tau}^{a}_{l}}}{\cal H}^{t}_1 \nonumber\\
&+&(-1)^{d_{{\tau}^{a}}}vQ^{-1}{\textstyle{\partial\over\partial
{\tau}^{a}_{l}}}
{\cal H}^{D}_2 -[(Dv)-v(Q^{-1}uv)]Q^{-1}
{\textstyle{\partial\over\partial {\tau}^{a}_{l}}}{\cal H}^{t}_1, \nonumber\\
{\textstyle{\partial\over\partial {\tau}^{a}_{l+1}}}u=
-{\textstyle{\partial\over\partial {\tau}^{a}_{l}}}u~'
&-& (-1)^{d_{{\tau}^{a}}}uQ^{-1}
{\textstyle{\partial\over\partial {\tau}^{a}_{l}}}
{\cal H}^{D}_2 -[(Du)+u(Q^{-1}uv)]Q^{-1}
{\textstyle{\partial\over\partial {\tau}^{a}_{l}}}{\cal H}^{t}_1 \nonumber\\
&-&(-1)^{d_{{\tau}^{a}}}uD^{-1}{\textstyle{\partial\over\partial
{\tau}^{a}_{l}}}{\cal H}^{Q}_2 -
[(Qu)-u(D^{-1}uv)]D^{-1}
{\textstyle{\partial\over\partial {\tau}^{a}_{l}}}{\cal H}^{t}_1 \nonumber\\
&+&2u[\theta,D]{\textstyle{\partial\over\partial
{\tau}^{a}_{l}}}{\cal H}^{t}_1,
\label{recrel}
\end{eqnarray}
where $d_{{\tau}^{a}}$ is the Grassmann parity of the evolution derivative
${\textstyle{\partial\over\partial {\tau}^{a}_{l}}}$ and
\begin{eqnarray}
{\cal H}^{t}_1\equiv  uv, \quad
{\cal H}^{D}_2\equiv  vDu + uvQ^{-1}(uv), \quad
{\cal H}^{Q}_2\equiv vQu - uvD^{-1}(uv)
\label{hamdens}
\end{eqnarray}
are the densities of the Hamiltonians ${H}^{t}_1$ \p{hams} as well as
${H}^{D}_2$ and ${H}^{Q}_2$ \p{hamsferm}, respectively.

Taking into account the involution properties
\begin{eqnarray}
&& \quad \quad \quad \quad \quad
-({\cal H}^{t}_1)^{*}=({\cal H}^{t}_1)^{\dagger}={\cal H}^{t}_1,
\nonumber\\
&&({\cal H}^{Q}_2)^{*}=- Q{\cal H}^{t}_1+{\cal H}^{Q}_2, \quad
({\cal H}^{D}_2)^{*}=- D{\cal H}^{t}_1+{\cal H}^{D}_2, \nonumber\\
&&({\cal H}^{Q}_2)^{\dagger}= D{\cal H}^{t}_1-{\cal H}^{D}_2, \quad
\quad ({\cal H}^{D}_2)^{\dagger}= Q{\cal H}^{t}_1-{\cal H}^{Q}_2
\label{conjf1}
\end{eqnarray}
of the Hamiltonian densities \p{hamdens},
one can verify that the recurrence relations (\ref{recrel}) possess
the involutions (\ref{conj*}--\ref{conjdagger}). Together with the already
verified fact that the first flows (\ref{eqs}--\ref{ueqs})
also admit these involutions one concludes that the all other flows of the
hierarchy under consideration admit them as well.

Using eqs. (\ref{recrel}) and \p{ff2-}, we obtain, for example, the third
bosonic flow
\begin{eqnarray}
{\textstyle{\partial\over\partial t_3}} v &=&
v~''' +3(Dv)~'(Quv)-3(Qv)~'(Duv)
+3v~'(Du)(Qv) \nonumber\\
&&-3v~'(Qu)(Dv)+6vv~'(DQu)-6(uv)^2v~', \nonumber\\
{\textstyle{\partial\over\partial t_3}} u &=&
u~''' +3(Qu)~'(Duv)-3(Du)~'(Quv)
+3u~'(Qv)(Du) \nonumber\\ &&-3u~'(Dv)(Qu)
+6uu~'(QDv)-6(uv)^2u~'
\label{flow3}
\end{eqnarray}
which coincides with the corresponding flow that can be derived from
the Lax-pair representation \p{qfl}. Let us underline that all $U_l$
flows for $l \geq 1$ can also be derived in this way starting from
the zero flow $U_0$ \p{ueqs} as an input.

Finally, let us transform the first bosonic and fermionic flows from eqs.
(\ref{eqs}--\ref{ueqs}) and recurrence relations \p{recrel} to the
basis \p{N2basis}, where they  become
\begin{eqnarray}
&&{\textstyle{\partial\over\partial t_1}}
\left(\begin{array}{cc} v\\ u \end{array}\right) =
{\partial}\left(\begin{array}{cc} v\\ u \end{array}\right), \quad
U_0\left(\begin{array}{cc} v\\ u \end{array}\right) =
-\frac{i}{\sqrt{2}}{\theta}{\cal D}
\left(\begin{array}{cc} v\\ u \end{array}\right),\nonumber\\
&&{\cal D}_1 v=-{\cal D}v-2v{\partial}^{-1}{\cal D}(uv),\quad
{\cal D}_1 u=-{\cal D}u+2u{\partial}^{-1}{\cal D}(uv),\nonumber\\
&&{\overline {\cal D}}_1 v=-{\overline {\cal D}}v+
2v{\partial}^{-1}{\overline {\cal D}}(uv),\quad
{\overline {\cal D}}_1 u=-{\overline {\cal D}}u-
2u{\partial}^{-1}{\overline {\cal D}}(uv),
\label{N2supersflows}
\end{eqnarray}
\vfill\eject
\begin{eqnarray}
{\textstyle{\partial\over\partial {\tau}^{a}_{l+1}}}v&=&
+{\textstyle{\partial\over\partial {\tau}^{a}_{l}}}v~' \nonumber\\
&+&(-1)^{d_{{\tau}^{a}}} [v{\partial}^{-1}{\cal D}
{\textstyle{\partial\over\partial {\tau}^{a}_{l}}}
(v{\overline {\cal D}}u-uv{\partial}^{-1}{\overline {\cal D}}uv)
-v{\partial}^{-1}{\overline {\cal D}}
{\textstyle{\partial\over\partial {\tau}^{a}_{l}}}
(v{\cal D}u+uv{\partial}^{-1}{\cal D}uv)] \nonumber\\
&-&[({\overline {\cal D}}v)+v({\partial}^{-1}
{\overline {\cal D}}uv)]{\partial}^{-1}{\cal D}
{\textstyle{\partial\over\partial {\tau}^{a}_{l}}}(uv)
+[({\cal D}v)-v({\partial}^{-1}{\cal D}uv)]{\partial}^{-1}
{\overline {\cal D}}
{\textstyle{\partial\over\partial {\tau}^{a}_{l}}}(uv),\nonumber\\
{\textstyle{\partial\over\partial {\tau}^{a}_{l+1}}}u&=&
-{\textstyle{\partial\over\partial {\tau}^{a}_{l}}}u~'\nonumber\\
&-&(-1)^{d_{{\tau}^{a}}}[u{\partial}^{-1}
{\overline {\cal D}}{\textstyle{\partial\over\partial {\tau}^{a}_{l}}}
(u{\cal D}v-uv{\partial}^{-1}{\cal D}uv)-
u{\partial}^{-1}{\cal D}{\textstyle{\partial\over\partial
{\tau}^{a}_{l}}}
(u{\overline {\cal D}}v+uv{\partial}^{-1}{\overline {\cal D}}uv)]
\nonumber\\
&+&[({\cal D}u)+u({\partial}^{-1}{\cal D}uv)]{\partial}^{-1}
{\overline {\cal D}}{\textstyle{\partial\over\partial
{\tau}^{a}_{l}}}(uv)
-[({\overline {\cal D}}u)-u({\partial}^{-1}{\overline {\cal D}}uv)]
{\partial}^{-1}{\cal D} {\textstyle{\partial\over\partial
{\tau}^{a}_{l}}}(uv).
\label{recrelN2basis}
\end{eqnarray}
These equations are obviously invariant under
the $GL(1)$ transformation \p{N2u1}. Consequently,
all higher flows admit this automorphism as well. Despite of this,
the Lax operator ${\cal M}$ \p{def1} is not invariant
with respect to the $GL(1)$ transformation. Hence, applying it to
${\cal M}$ one can derive a one-parameter family of consistent Lax
operators,
\begin{eqnarray}
\quad \quad \quad \quad
{\cal M} \quad \Longrightarrow \quad {\cal M}^{\phi}=
\cos \phi ~{\cal M}+ \sin \phi ~({\cal M}^{\dagger})^{T}
\label{ulax}
\end{eqnarray}
with ${\cal M}^{\dagger}$ defined in eq. \p{conjlax}. The flows generated
in this way are all isomorphic. We remark that the superfields $v$ and
$u$ have trivial transformation properties under the $GL(1)$
transformation \p{N2u1}, while the superfunctions $w_n^{f}$ and $w_n^{b}$
\p{def11} expressed in terms of these transform in a rather complicated
manner.

\section{Secondary reduction: a new N=2 supersymmetric \\
modified KdV hierarchy}
In this section we derive a new $N=2$ supersymmetric modified KdV
hierarchy by means of the secondary reduction.

Let us investigate the secondary reduction of the hierarchy
considered in the preceding sections. We impose the
following secondary constraint\footnote{See also
refs. \cite{yu,rs,ra}, where a similar reduction of the Manin-Radul
\cite{mr} and Mulase-Rabin \cite{m,r} $N=1$ supersymmetric KP and KdV
hierarchies has been discussed.} on the Lax operator ${\cal L}$ \p{def1}:
\begin{eqnarray}
{\cal M}^{T} = D{\cal M}D^{-1}
\label{red}
\end{eqnarray}
which can easily be resolved in terms of the superfield $v$
entering ${\cal M}$,
\begin{eqnarray}
v=1.
\label{red1}
\end{eqnarray}
Then, the reduced Lax operator ${\cal M}^{red}$ becomes
\begin{eqnarray}
{\cal M}^{red}= Q +  D^{-1} u.
\label{red3}
\end{eqnarray}
Condition \p{red} by means of eq. \p{def1eq} induces the secondary
constraint
\begin{eqnarray}
({{\cal W}^{-1}})^{T} = D{\cal W}D^{-1}
\label{red4}
\end{eqnarray}
on the dressing operator ${\cal W}$ (\ref{def11})
which in turn induces the following secondary constraints on the operators
$L_l$ \p{laxkp}:
\begin{eqnarray}
&&({\cal L}_{2l})^{T} = (-1)^{l} D{\cal L}_{2l}D^{-1}, \quad
({\cal L}_{2l-1})^{T} = (-1)^{l} D{\cal L}_{2l-1}D^{-1}
\label{red5}
\end{eqnarray}
which are identically satisfied if constraint \p{red} (or \p{red4})
is imposed. Importantly, eqs. \p{red5} imply that
\begin{eqnarray}
({\cal L}_{2(2k-1)})_{0} = ({\cal L}_{2(2k)-1})_{0} =0, \quad k= 1,2 \ldots ,
\label{red6}
\end{eqnarray}
where the subscript $0$ refers to the constant part of the operators.
Consequently, the equations:
\begin{eqnarray}
(({\cal L}_{2(2k-1)})_{+}1) = (({\cal L}_{2(2k)-1})_{+}1) = 0
\label{red7}
\end{eqnarray}
are identically satisfied as well.
Using these relations, the involution \p{conjdagger}, and the algebraic
structure (\ref{alg1}--\ref{algqqbar}) we are led to the conclusion that
only half of the flows (\ref{modflv}) are consistent with the reduction
(\ref{red}--\ref{red1}), namely
\begin{eqnarray}
\{~{\textstyle{\partial\over\partial t_{2k-1}}},
~U_{2k},~D_{2k},~Q_{2k}~\}.
\label{consistfl}
\end{eqnarray}

In order to understand better what kind of reduced hierarchy we have
in fact derived, one might analyze its Hamiltonian structure via
Hamiltonian reduction of the first and second Hamiltonian structures
(\ref{hamstr}--\ref{hamstr2}) we started with.
However, it is easier to reduce the less complicated expressions
\p{hamstr2bf}. In this basis, the constraint \p{red1} becomes
\begin{eqnarray}
f=0,
\label{red2}
\end{eqnarray}
and the superfield $b$ coincides with the superfield $u$
on the constraint surface.

Let us start with the first Hamiltonian structure $J^{(b,f)}_{1}$
\p{hamstr2bf}. In this case, the constraint \p{red2} is a gauge
constraint, and a gauge can be fixed by the condition $b=0$. As the
result, the trivial reduced Hamiltonian structure is generated.

In the case of the second Hamiltonian structure
$J^{(b,f)}_{2}$ \p{hamstr2bf}, the constraint \p{red2} is second
class, and we can use Dirac brackets in order to obtain
the second Hamiltonian structure for the reduced system. The result is
\begin{eqnarray}
{J^{(Dirac)}_{11}}=J^{(u,0)}_{11}-
J^{(u,0)}_{12}{J^{(u,0)}_{22}}^{-1}J^{(u,0)}_{21}\equiv
\frac{1}{2}({\partial}DQ+QuQ-DuD
+2{\partial} u + 2u {\partial}), ~ ~ ~
\label{redhamstr2bf}
\end{eqnarray}
where we have exploited the relations
\begin{eqnarray}
J^{(b,0)}_{12}Q=\frac{1}{2}
QJ^{(b,0)}_{22}Q=QJ^{(b,0)}_{21}= -{\partial}DQ-QbQ+DbD
\label{red1hamstr2bf}
\end{eqnarray}
which can easily be read off eqs. \p{hamstr2bf}.

From eqs. \p{redhamstr2bf} we see that the second
Hamiltonian structure of the secondary reduced hierarchy displays
the reduced $N=2$ superconformal structure, and its flows
\p{consistfl} possess a global $N=2$ supersymmetry with an unusual
length dimensions of its generators,
\begin{eqnarray}
[{\textstyle{\partial\over\partial t_3}}]=-3, \quad
[U_0]=0, \quad [D_2]=[Q_2]=-\frac{3}{2}.
\label{n2supred}
\end{eqnarray}
Substituting the constraint \p{red1} into the third flow
equations \p{flow3} of the reduced hierarchy, they become
\begin{eqnarray}
{\textstyle{\partial\over\partial t_3}} u &=&
(u~'' -3(Du)(Qu)+2u^3)~',
\label{redflow3}
\end{eqnarray}
and one can easily recognize that this equation reproduces
the modified KdV equation in the bosonic limit when the fermionic
component is put equal to zero. Equation \p{redflow3} does not
coincide with any of the three known $N=2$ extensions \cite{lm} of the
modified KdV equation. Therefore, we summarize that the secondary reduced
hierarchy gives a new type of $N=2$ supersymmetric generalization of
the modified KdV hierarchy.

\section{Generalizations, Conclusion and Outlook}
In this section we discuss possible generalizations of the reduced $N=2$
KP hierarchy to the matrix case and some open problems.

The hierarchies discussed in the preceding sections admit a natural
generalization to the non-abelian case. One may
consider the $N=2$ supersymmetric matrix KP hierarchy generated by a
matrix-valued dressing operator $W$ in $N=1$ superspace,
\begin{eqnarray}
W \equiv I+\sum_{n=1}^\infty ~(w^{(b)}_n +w^{(f)}_nD)~{\partial}^{-n},
\label{matrixkp}
\end{eqnarray}
which can be treated as a reduction of the analogous operator
in $N=2$ superspace considered in \cite{dgs}. Its consistent
reductions are characterized by the reduced operator
\begin{eqnarray}
{\cal M}_{1}= IQ + vD^{-1}u.
\label{matrixred}
\end{eqnarray}
Here, $w_n\equiv (w_n)_{AB}(Z)$, $v\equiv v_{Aa}(Z)$ and $u\equiv
u_{aA}(Z)$
($A,B=1,\ldots, k$; $a,b=1,\ldots , n+m$) are
rectangular matrix-valued superfields,
and $I$ is the identity matrix, $I\equiv {\delta}_{A,B}$.
In \p{matrixred} the matrix product is understood, for
example $(vu)_{AB} \equiv \sum_{a=1}^{n+m} v_{Aa}u_{aB}$.
The matrix entries are bosonic superfields for $a=1,\ldots ,n$ and
fermionic superfields for $a=n+1,\ldots , n+m$, i.e.,
$v_{Aa}u_{bB}=(-1)^{d_{a}{\overline d}_{b}}u_{bB}v_{Aa}$, where $d_{a}$
and ${\overline d}_{b}$ are the Grassmann
parities of the matrix elements $v_{Aa}$ and $u_{bB}$,
respectively, $d_{a}=1$ $(d_{a}=0)$ for fermionic (bosonic) entries.
The grading choosen guarantees that the Lax operator
${\cal M}_{1}$ is Grassmann odd \cite{bks2}.

A detailed analysis of the emerging hierarchies is, however, beyond the
scope of the present paper. Without going into more details, let us only
present a few non-trivial bosonic and fermionic
flows in this noncommutative case (compare with the abelian flows
(\ref{eqs}--\ref{ff2-}) ):
\begin{eqnarray}
{\textstyle{\partial\over\partial t_2}} v =
+v~'' -  2\{Q,vDu\}v-2v(uv)^2, & \quad &
{\textstyle{\partial\over\partial t_2}} u =
-u~'' +  2\{D,uQv\}u+2(uv)^2u, \nonumber\\
D_1 v= -Dv+2(Q^{-1}v{\cal I}u)v, & \quad &
D_1 u=-Du-2uQ^{-1}(vu), \nonumber\\
Q_1 v= -Qv-2vD^{-1}(uv), & \quad &
Q_1 u= -Qu+2{\cal I} (D^{-1}uv)u
\label{eqsmatrix}
\end{eqnarray}
which are derived using Lax-pair representations (\ref{qfl})
with ${\cal M}_1$ \p{matrixred} and
\begin{eqnarray}
({\cal L}_1)_{+} = ID-2(Q^{-1}(v{\cal I}u)),
\label{Mmatrix}
\end{eqnarray}
and the matrix ${\cal I}$ is defined as
\begin{eqnarray}
{\cal I} \equiv (-1)^{d_a} {\delta}_{ab}.
\label{matrI}
\end{eqnarray}
It is crucial that the existence of these two different fermionic first
flows, $D_1$ and $Q_1$ \p{eqsmatrix}, guarantees the $N=2$ supersymmetry
of the corresponding hierarchies.

For the particular case when the index $A$ takes only the value $A=1$,
the matrix reduced Lax operator \p{matrixred} becomes a scalar
operator generating a reduced hierarchy with $n+m$ pairs
of scalar superfields $v_a,u_a$. In the more special case
$A=1,a=1$ and $n=1, m=0$, the Lax operator \p{matrixred}
reproduces the Lax operator \p{def1}.

The results described in the previous sections can also be
generalized to the case of some other known reductions
of the supersymmetric KP hierarchy in $N=1$ superspace. For
example, a wide class of the following reductions
\begin{eqnarray}
L_{1} = {\cal L}\equiv  D + \sum_{a=1}^{m} v_a D^{-1} u_a, \quad m \in
\hbox{\bbd N}
\label{genred}
\end{eqnarray}
was proposed in \cite{anp1}. It is quite obvious that one can
generalize them by replacing the superfunctions $v_a$ and $u_a$ by
supermatrices with the above--described grading. A less
obvious fact is that one can consistently extend in an $N=2$
supersymmetric fashion the number of
bosonic and fermionic flows of the reduced hierarchies obtained in
\cite{anp1}. To simplify the consideration let us concentrate on the
simplest example of the reduced hierarchy, characterized by the Lax
operator \p{genred} at $m=1$. Then, as it was
shown in \cite{anp1}, the Lax operator ${\cal L}$ \p{genred} satisfies
an equation which can be read from eq. \p{ar} by replacing the
operator ${\cal M}$ by ${\cal L}$ there, and the flows
\begin{eqnarray}
 {\textstyle{\partial\over\partial t_l}}{\cal W}=
-({\cal L}_{2l})_-{\cal W}, \quad
D_l{\cal W}=-(({\cal L}_{2l-1})_- - {\widetilde {\cal L}}_{2l-1}){\cal W}
\label{satoflowred}
\end{eqnarray}
can consistently be introduced. Here, the operator ${\widetilde {\cal 
L}}_{2l-1}$ can be read off eq. \p{conj3} by replacing the operators 
${\widetilde {\cal M}}_{2l-1}$ and ${\cal M}$ there by the operators 
${\widetilde {\cal L}}_{2l-1}$ and ${\cal L}$, respectively.  Now, one can 
easily observe that the operators ${\cal L}$ \p{satoflowred} and ${\cal M}$ 
\p{def1} possess the same properties in spite of their different appearance, 
and for this reason one can construct the same set of consistent Sato 
equations \p{satoflow} for each of them. Comparing equations \p{satoflowred} 
with \p{satoflow} shows that at least one more series of fermionic flows, 
namely 
\begin{eqnarray} 
Q_l {\cal W}= -({\cal M}_{2l-1})_-{\cal W} 
\label{satoflowredq}
\end{eqnarray}
can consistently be added to the Sato equations \p{satoflowred}
and, consequently, the extended hierarchy of the flows are indeed $N=2$
supersymmetric.

The hierarchies proposed in this paper may
appear to have come out of the blue. It is time to explain how we were
lead to their construction by relating them to previously known
hierarchies. Forerunners of the present paper are refs.
\cite{tak,ls,ols} and especially refs. \cite{anp1,dgs}. As one might
suspect, there is a correspondence between the $N=2$ supersymmetric
hierarchies defined above and the $N=4$ supersymmetric hierarchies
proposed in \cite{dgs}, but this correspondence is rather
non-trivial and indirect. The heuristic analysis of the $N=4$ flows
constructed in \cite{dgs} shows that among them exist flows which
contain only the operators $D_{+}$ and $D_{-}$ (and not $Q_{+}$ and
$Q_{-}$) and which are in some sense $N=2$ like. Restricting
the whole hierarchy to only these flows, one can consistently
reduce them by the constraint ${\theta}_{+}=i{\theta}_{-}\equiv {\theta}$ 
which leads to the correspondence $D_+\equiv D$ and $D_-\equiv iQ$ with the 
fermionic derivatives of the present paper, where $i$ is the imaginary unity 
and ${\theta}_{\pm}$ are the Grassmann coordinates of $N=2$ superspace. This 
constraint is consistent for the algebra of the fermionic derivatives 
$D_{\pm}$, but it is surely inconsistent for the algebra extended by 
any of the two fermionic derivatives $Q_{\pm}$. Without going into details 
we would like to stress that this reduction is a rather 
nontrivial one, and the whole construction given in \cite{dgs} must  
properly be adjusted. For illustrative purposes consider
the product $D_+D_{-}^{-1}$ which appears
when constructing the consistent Sato equations. This 
product has no differential piece before the reduction, 
but it becomes a purely {\it differential(!)} operator by virtue of the 
reduction constraint, drastically changing the construction.
Moreover, our $U_l$ flows cannot be
derived by reducing the $N=4$ flows, so they must be added by hand
in order to complete the hierarchy.
To close this discussion let us state two unsolved
questions whose answering should yield a deeper
understanding of the proposed hierarchies:

1. What is the consistent Lax-pair representation of the $U_l$ flows?

2. What are proper general formulae for the Hamiltonians
$H^{U}_l$, $H^{D}_l$ and $H^{Q}_l$ analogous to formula \p{res0}
for the Hamiltonian $H^{t}_l$?

We hope to return to this questions elsewhere.

Finally, we would like to briefly comment on some unusual properties of our 
hierarchy.

1. Our hierarchy flows in $N=1$ superspace contain both the $N=1$
fermionic derivative $D$ and the $N=1$ supersymmetry generator $Q$,
nevertheless they are $N=2$ supersymmetric. The resolution of this sophism
is hidden in the nonlocal character of the $N=2$ transformations.

2. The equations for the bosonic components of our bosonic
flows ${\textstyle{\partial\over\partial t_l}}$ do not contain the
fermionic components at all. Nevertheless, the supersymmetrization of these
equations is non-trivial\footnote{By trivial supersymmetrization of 
bosonic equations we mean just replacing functions by superfunctions.
In this case the resulting equations are $N=2$ supersymmetric as well, 
but they do not contain the fermionic derivatives at all.} because it 
involves the fermionic operators $D$ and $Q$.  

3. The residue \p{defres} we used for pseudo-differential operators in
$N=1$ superspace is not the usual $N=1$ residue which is the
coefficient of the operator $D^{-1}$. We obtained this unusual definition
for the residue by the above--explained reduction of the residue
introduced in \cite{dgs}.

4. Grassmann-odd Hamiltonian structures appear at the Hamiltonian
description of our supersymmetric hierarchy.
To our knowledge, this is the first example of a non-trivial
supersymmetrized hierarchy with odd bi-Hamiltonian structure.
It is interesting to speculate whether an even
bi-Hamiltonian structure exists as well.

5. The secondary reduced hierarchy is a new $N=2$ supersymmetric
modified KdV hierarchy with unusual length dimensions of the
$N=2$ supersymmetry generators (see, eqs. \p{n2supred}).

All these peculiarities once more demonstrate the rich structure encoded
in supersymmetry.

{}~

{}~

\noindent{\bf Acknowledgments.}
A.S. would like to thank F. Delduc for useful discussions and the Institut
f\"ur Theoretische Physik, Universit\"at Hannover for the financial
support and hospitality during the course of this work. This work was
partially supported by the Heisenberg-Landau programme HLP-99-13,
PICS Project No. 593, RFBR-CNRS Grant No. 98-02-22034, RFBR Grant No.
99-02-18417, INTAS Grant INTAS-96-0538 and Nato Grant No. PST.CLG 974874.

\vfill\eject

\section*{Appendix.
A new N=2 supersymmetric Toda chain}
\setcounter{equation}{0}
\def\theequation{A.\arabic{equation}}
In this appendix we present a new version
of the supersymmetric Toda chain equation and derive its zero-curvature
representation which lies at the origin of the reduction constraint
\p{def1}.

Let us introduce the new equation
\begin{eqnarray}
QD\ln b_i=b_{i+1}-b_{i-1}
\label{zerocurveqs2}
\end{eqnarray}
written in terms of the bosonic $N=1$ superfields $b_i\equiv
b_i(z,\theta)$ defined on the chain, $i\in \hbox{\bbd Z}$.
This equation represents a one-dimensional $N=1$
generalization of the
two-dimensional $N=(1|1)$ superconformal Toda lattice equation.
It can be rewritten as a system of two equations
\begin{eqnarray}
Qf_i=b_i+b_{i+1}, \quad D\ln b_i=f_i-f_{i-1}
\label{zerocurveqs1}
\end{eqnarray}
which admits the zero-curvature representation
\begin{eqnarray}
\{D-A^{\theta}_{D}~,~Q-A^{\theta}_{Q}\}=0
\label{zerocurv}
\end{eqnarray}
with the fermionic connections
\begin{eqnarray}
(A^{\theta}_{D})_{ij}\equiv f_i{\delta}_{i,j}+{\delta}_{i,j-1}, \quad
(A^{\theta}_{Q})_{ij} \equiv -b_i{\delta}_{i,j+1},
\label{fermconn}
\end{eqnarray}
where $f_i\equiv f_i(z,\theta)$
are fermionic $N=1$ chain superfields. One can
define the bosonic connections $A_{z^{\pm}}$ by
\begin{eqnarray}
{\partial} +A^{z}_{D}\equiv (D-A^{\theta}_{D})^2, \quad
{\partial} +A^{z}_{Q}\equiv -(Q-A^{\theta}_{Q})^2.
\label{boscon}
\end{eqnarray}
More explicitly, they read
\begin{eqnarray}
(A^{z}_{Q})_{ij}\equiv -Qb_i{\delta}_{i,j+1} +b_ib_{i-1}{\delta}_{i,j+2},
\quad (A^{z}_{D})_{ij}\equiv -D_+f_i{\delta}_{i,j}+
(f_i-f_{i+1}){\delta}_{i,j-1} - {\delta}_{i,j-2}
\label{connect}
\end{eqnarray}
an,d due to \p{zerocurv}, obviously satisfy the zero-curvature condition
\begin{eqnarray}
[{\partial} + A^{z}_{Q}~,~{\partial}+A^{z}_{D}]=0
\label{bozcon1}
\end{eqnarray}
which is a consistency condition for the linear system
\begin{eqnarray}
\quad ~({\partial}+A^{z}_{Q})\Psi=\lambda \Psi, \quad
\label{linsys1}
\end{eqnarray}
\begin{eqnarray}
({\partial}+A^{z}_{D})\Psi=0,
\label{linsys2}
\end{eqnarray}
where $\Psi\equiv \Psi_i$ is the chain wave function and $\lambda$ is a
spectral parameter. Taking into account the first relation of eqs.
\p{boscon}, equation \p{linsys2} can equivalently be rewritten in the
form
\begin{eqnarray}
(D -A^{\theta}_{D})\Psi=0.
\label{linsys3}
\end{eqnarray}
The linear system \p{linsys1}, \p{linsys3} is a key object in our
consideration.

In order to derive
the Lax operator we are looking for, we follow a trick
proposed in \cite{bx} and express each chain function entering the
spectral equation \p{linsys1} in terms of chain functions defined at
the single chain point $i$, using eqs. \p{zerocurveqs1}
and \p{linsys3}. In this manner 
we obtain the new spectral equation
\begin{eqnarray}
(Q + \frac{1}{D-f_i} b_i)^2\Psi_i=\lambda\Psi_i.
\label{lax1}
\end{eqnarray}
For each fixed value of $i$, it represents the spectral equation of
the differential hierarchy, i.e. of the hierarchy of equations involving
only the superfields $b_i,f_i$ at a single lattice point. Applying the
discrete chain shift (i.e., the system of eqs. \p{zerocurveqs1}) 
to the differential hierarchy generates the discrete
hierarchy. Thus, the discrete hierarchy appears as a
collection of an infinite number of isomorphic differential hierarchies
\cite{bx}.

It is well known that a spectral equation is just an equation for a Lax
operator. For a fixed value of $i$ one can completely omit the chain index
in the spectral equation \p{lax1}, and it is obvious that the operator
\begin{eqnarray}
M=(Q + \frac{1}{D-f} b)^2
\label{lax2}
\end{eqnarray}
is just the Lax operator which is responsible for the bosonic flows of the
differential hierarchy. In the new superfield basis
$\{v_i,u_i\}$ defined by
\begin{eqnarray}
b_i\equiv u_iv_i, \quad f_i\equiv D\ln v_i,
\label{zerocurveqs3}
\end{eqnarray}
in which the system \p{zerocurveqs1} becomes an $N=1$
supersymmetric generalization of the Darboux transformation \p{discrsymm}
\begin{eqnarray}
u_{i+1}=\frac{1}{v_{i}}, \quad
QD\ln v_i=u_{i+1}v_{i+1}+u_{i}v_{i},
\label{Darboux}
\end{eqnarray}
the Lax operator \p{lax2} simplifies to
\begin{eqnarray}
M=(Q + v D^{-1} u)^2.
\label{sollin}
\end{eqnarray}
Let us remark that the operator ${\cal M}$ \p{def1}
is just the square root of the operator $M$ \p{sollin}.
This Lax operator has been used in section 3
for constructing a consistent reduction of all other flows of the
$N=2$ supersymmetric KP hierarchy.

\vfill\eject


\begin{thebibliography}{**}

\bibitem{mr}
Yu.I. Manin and A.O. Radul, {\it A supersymmetric extension of the
Kadomtsev-Petviashvili hierarchy}, Commun. Math. Phys. {\bf 98} (1985) 65.
\bibitem{m}
M. Mulase, {\it A new super KP system and a characterization
of the Jacobians of arbitrary algebraic super curves},
J. Diff. Geom. {\bf 34} (1991) 651.
\bibitem{r} J. Rabin, {\it The geometry of the Super KP flows},
Commun. Math. Phys. {\bf 137} (1991) 552.
\bibitem{tak}
M. Takama, {\it Grassmannian approach to Super--KP hierarchies},\\
YITP/U-95-23, hep-th/9506165.
\bibitem{dgs}
F. Delduc, L. Gallot and A. Sorin,
{\it N=2 local and N=4 nonlocal reductions of supersymmetric KP
hierarchy in N=2 superspace}, LPENSL-TH-14/99, solv-int/9907004,\\
Nucl. Phys. {\bf B}, to appear.
\bibitem{anp1}
H. Aratyn, E. Nissimov and S. Pacheva,\\
{\it Supersymmetric KP hierarchy: ``ghost'' symmetry structure,
     reductions and Darboux-B\"acklund solutions},
J. Math. Phys. {\bf 40} (1999) 2922, solv-int/9801021;\\
{\it Berezinian construction of super-solitons in supersymmetric
     constrained KP hierarchies}, in ``{\sl Topics in Theoretical Physics
vol. II}'' {\em Festschrift for A.H. Zimerman}, IFT-S\~{a}o Paulo,
SP-1998, pgs. 17-24, solv-int/9808004.
\bibitem{ls}
A.N. Leznov and A.S. Sorin,
{\it Two-dimensional superintegrable mappings and integrable
hierarchies in the $(2|2)$ superspace},
{\sl Phys. Lett.} {\bf B389} (1996) 494, hep-th/9608166;\\
{\it Integrable mappings and hierarchies in the $(2|2)$ superspace},\\
{\sl Nucl. Phys.} (Proc. Suppl.) {\bf B56} (1997) 258.
\bibitem{ols}
O. Lechtenfeld and A. Sorin,
{\it Fermionic flows and tau function of the N=$(1|1)$ superconformal Toda
lattice hierarchy}, ITP-UH-23/98, JINR E2-98-285, solv-int/9810009,\\
Nucl. Phys. {\bf B}, to appear.
\bibitem{stan}
S. Stanciu, {\it Additional symmetries of supersymmetric KP
hierarchies},\\
Commun. Math. Phys. {\bf 165} (1994) 261, hep-th/9309058.
\bibitem{rad}
A.O. Radul,
Sov. Phys. JETP Lett. {\bf 50} (1989) 371.
\bibitem{eor}
B. Enriquez, A.Yu. Orlov and V.N. Rubtsov, {\it Dispersionful analogues
of Benney's equations and N-wave systems}, Inverse Problems {\bf 12}
(1996) 241, solv-int/9510002.
\bibitem{ffs}
J.M. Figueroa-O'Farrill and S. Stanciu,
{\it On a new supersymmetric KdV hierarchy in 2-d quantum supergravity},
Phys. Lett. {\bf B316} (1993) 282, hep-th/9302057;\\
{\it New supersymmetrization of
the generalized KdV hierarchies},\\ Mod. Phys. Lett. {\bf A8} (1993) 2125,
hep-th/9303168.
\bibitem{mcar}
I.N. McArthur, {\it Odd Poisson brackets and the fermionic hierarchy of
Becker and Becker}, J. Phys. {\bf A26} (1993) 6379.
\bibitem{bd}
J.C. Brunelli and A. Das, {\it The sTB-B hierarhy}, \\
Phys. Lett. {\bf B409} (1997) 229, hep-th/9704126.
\bibitem{s}
V. A. Soroka, {\it Linear odd Poisson bracket on Grassmann variables}, \\
Phys. Lett. {\bf B451} (1999) 349, hep-th/9811252. 
\bibitem{p}
Z. Popowicz,
{\it Odd bihamiltonian structure of new supersymmetric N=2,4 Korteweg de
Vries equation and odd SUSY Virasoro-like algebra}, hep-th/9903198.
\bibitem{ff} 
B. Fuchssteiner and A.S. Fokas,
{\it Symplectic structures, their B\"acklund transformations
and hereditary symmetries}, Physica {\bf D4} (1981) 47.
\bibitem{yu}
F. Yu, {\it Bi-Hamiltonian structure of super KP hierarchy},
hep-th/9109009.
\bibitem{rs}
E. Ramos and S. Stanciu, {\it On the supersymmetric BKP hierarchy},\\
Nucl. Phys. {\bf B427} (1994) 338, hep-th/9402056.
\bibitem{ra}
E. Ramos, {\it A comment on the odd flows for the supersymmetric KdV
hierarchy}, \\ Mod. Phys. Lett. {\bf A9} (1994) 3235, hep-th/9403043.
\bibitem{lm}
P. Laberge and P. Mathieu, {\it N=2 superconformal algebra and integrable
O(2) fermionic extensions of the Korteweg-de Vries equation},
Phys. lett. {\bf B215} (1988) 718;\\
P.Labelle and P. Mathieu, {\it A new N=2 supersymmetric Korteweg-de Vries
equation},\\ J. Math. Phys. {\bf 32} (1991) 923.
\bibitem{bks2}
L. Bonora, S. Krivonos and A. Sorin, {\it The N=2 supersymmetric matrix
GNLS hierarchies}, Lett. Math. Phys. {\bf 45} (1998) 63, solv-int/9711009.
\bibitem{bx}
L. Bonora and C.S. Xiong,
{\it An alternative approach to KP hierarchy in matrix models},
{\sl Phys. Lett.} {\bf B285} (1992) 191, hep-th/9204019;\\
{\it Matrix models without scaling limit},
{\sl Int. J. Mod. Phys.} {\bf A8} (1993) 2973, hep-th/9209041.

\end{thebibliography}
\end{document}